\documentclass{iaus}
\usepackage{graphicx}
\usepackage{subfigure}

\def\apj{\textit{ApJ}}
\def\aap{\textit{A\&A}}
\def\mnras{\textit{MNRAS}}
\def\araa{\textit{ARAA}}

\def\msol{ M_\odot}
\def\mjup{ M_{\rm J}}
\def\rjup{R_{\rm J}}
\def\mearth{\,{ M}_\oplus}

\def\beq{\begin{equation}}
\def\eeq{\end{equation}}
\def\simgr{\,\hbox{\hbox{$ > $}\kern -0.8em \lower 1.0ex\hbox{$\sim$}}\,}
\def\simle{\,\hbox{\hbox{$ < $}\kern -0.8em \lower 1.0ex\hbox{$\sim$}}\,}

\title[Understanding exoplanets ] 
{Understanding exoplanet formation, structure and evolution in 2010}

\author[G. Chabrier]  
{Gilles Chabrier$^{1,2}$, J\'er\'emy Leconte$^{1}$ \&  Isabelle Baraffe$^{2,1}$ }

\affiliation{$^1$  \'Ecole Normale Sup\'erieure de Lyon, CRAL (CNRS, UMR 5574), F-69364 Lyon cedex 07, France
  \\[\affilskip]
 $^2$
Physics \& Astronomy, University of Exeter, Exeter EX4 4PE, UK\\
(chabrier, jeremy.leconte, ibaraffe@ens-lyon.fr)
}

\pubyear{2010}
\volume{276}  
\pagerange{TBD}
\jname{The Astrophysics of Planetary Systems: Formation, Structure, and Dynamical Evolution}
\editors{A. Sozzetti, M. G. Lattanzi and A. P. Boss, eds.}

\begin{document}

\maketitle

\begin{abstract}
In this short review, we summarize our present understanding (and non-understanding) of exoplanet formation, structure and evolution, in the light of the most recent discoveries. Recent observations of transiting massive brown dwarfs seem to remarkably confirm the predicted theoretical mass-radius relationship in this domain. This mass-radius relationship provides, in some cases, a powerful diagnostic to distinguish planets from brown dwarfs of same mass, as for instance for Hat-P-20b. If confirmed, this latter observation shows that planet formation takes place up to at least 8 Jupiter masses. Conversely, observations of brown dwarfs down to a few Jupiter masses in young, low-extinction clusters strongly suggests an overlapping mass domain between (massive) planets and (low-mass) brown dwarfs, i.e. no mass edge between these two distinct (in terms of formation mechanism) populations. At last, the large fraction of heavy material inferred for many of the transiting planets confirms the core-accretion scenario as been the dominant one for planet formation.

\end{abstract}

\section{Planet internal structure and evolution}

\subsection{General overview}

The realm of extrasolar planet discoveries now extends from gaseous giants of several Jupiter masses down
 to objects
of a few Earth masses. Detailed
models of planet structure and evolution have been computed by different groups (Fortney et al. 2007, Baraffe et al. 2008, Burrows et al. 2007, Leconte et al. 2009; see Baraffe et al. 2010 for a recent review). These calculations include various internal compositions, based on presently available high-pressure equations of state (EOS) for materials typical of planetary interiors. A detailed discussion
and a comparison of these
models can be found in Baraffe et al. (2008)\footnote{Models are available at http://perso.ens-lyon.fr/isabelle.baraffe/PLANET08/}. This latter paper also explores the effect of the location
of the heavy element material in the planet, either all gathered at depth as a central core or distributed throughout the gaseous H/He envelope, on the
planet's radius evolution. These different
possible distributions of heavy elements can in some cases have an important impact on the planet's contraction.
This paper also shows
that the presence of even a modest gaseous (H/He) atmosphere hampers an accurate determination
of the planet's internal composition, as the highly compressible gas contains most of the entropy of the planet and thus governs its
cooling and contraction rate. In such cases, only the average internal composition of the planet
can be inferred from a comparison of the models with the observed mass and radius determinations, for transiting objects. 

Objects below about 10 Earth-masses, globally denominated Super-Earth or Earth-like planets, on the other hand, are not massive enough to retain a significant gaseous atmosphere by gravitational instability (Mizuno 1980, Stevenson 1982, Rafikov 2006). For these objects, the lack of a substantial gaseous atmosphere allows a more
detailed exploration of the planet's internal composition than for the gaseous planets (Valencia et al. 2007, Seager et al. 2007, Sotin et al. 2007).
It should be kept in mind, however, that, even for these Super-Earth or Earth-like planets, present uncertainties in the EOS of the various heavy elements (e.g. H$_2$O, Fe) under relevant P-T interior conditions prevent an accurate determination of their internal composition (see e.g. Fig. 2 of  Baraffe et al. 2008). The melting lines of water or Iron are not even known under such conditions, so the exact thermodynamics state of these elements is unknown. The situation should improve in the coming years with the advent of high-pressure experiments conducted with the high-power laser facilities developed in the US and in France.

\begin{figure}[htbp] 
 \centering
 \resizebox{.7\hsize}{!}{\includegraphics{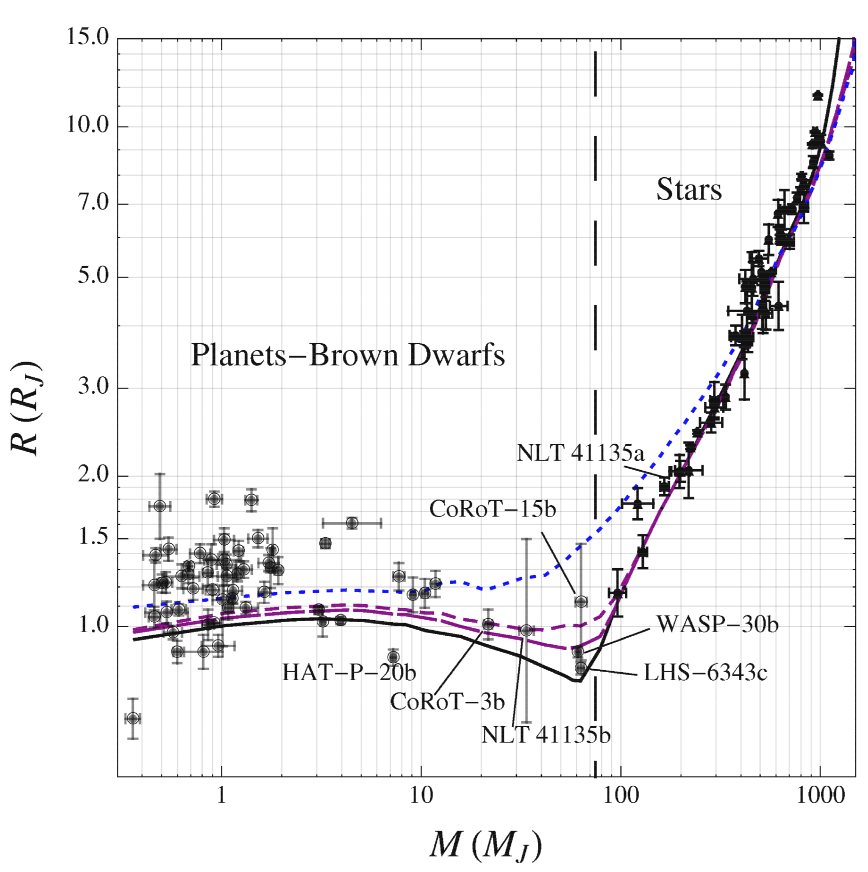}}
 \caption{Mass-radius relationship  from the stellar to the planetary regime, from one solar mass to one Saturn mass. The four curves display four isochrones, namely, from top to bottom, $10^8$ (dot), $5\times 10^8$ (short-dash), $10^9$ (long-dash) and $5\times 10^9$ (solid) yr. Some objects are identified on the figure, including the recent field M-dwarf/BD system NLTT 41135a,b (Irwin et al. 2010). The group of top 4 objects at $\sim 10\mjup$ includes Wasp\,14b, Hat-P-2b, Wasp\,18b and XO\,3b.  }
 \label{fig:MR}
\end{figure}

Figure \ref{fig:MR} displays the overall mass-radius relationship in the stellar and substellar domains, from a solar mass down to a Saturn mass. The lines denote the low-mass star, brown dwarf and planet models of the Lyon group for 4 isochrones. The vertical dash-line corresponds to the mass limit to reach thermal equilibrium, i.e. balance between nuclear H-burning energy and gravitational contraction energy, $M_{\mathrm{HBMM}}=0.075\,\msol$ (Chabrier \&  Baraffe 1997). This defines the limit between the stellar and brown dwarf domains. The general behaviour of this m-R relationship is discussed in detail in Chabrier \&  Baraffe (2000) and Chabrier et al. (2009) and will not be repeated here, where we will focus on the brown dwarf and planetary part of the domain. It is important however, to point out the recent observations of the $61\,\mjup$ and $63\,\mjup$ transiting brown dwarfs WASP-30b (Anderson et al. 2010) and LHS 6343 C (Johnson et al. 2010),
respectively, which remarkably confirm the predicted theoretical mass-radius relation in the brown dwarf domain (Chabrier \& Baraffe 2000).

As seen in the figure and explored in detail in Baraffe et al. (2008) and Leconte et al. (2009), for several of these transiting planets, the observed mass-radius relation can be adequately explained by the planet "standard" evolution models mentioned in \S1.1. A typical case, for instance, is CoRoT-Exo-4b, a 0.72$\mjup$ planet whose 1.17 $\rjup$ radius is reproduced at the 1$\sigma$ level by a model including a 10$\mearth$ water core surrounded by a gaseous H/He envelope, i.e. a global $\sim 5\%$ mass fraction of heavy material, more than twice the solar value (Leconte et al. 2009)\footnote{Models are available at:\\ http://perso.ens-lyon.fr/jeremy.leconte/JLSite/JLsite/Exoplanets\_Simulations.html}. Choosing rock or a mixture of water and rock instead of water as the main component of the core only slightly
changes this value. 
Several other transiting planet mass-radius signatures are well explained by standard models with moderate to high (up to $\sim 95\%$ for Neptune-mass planets) heavy element enrichment, as expected from the standard "core accretion" scenario for planet formation (see \S\ref{sec:formation}). 

\begin{figure}[htbp] 
 \centering
 \resizebox{.7\hsize}{!}{\includegraphics{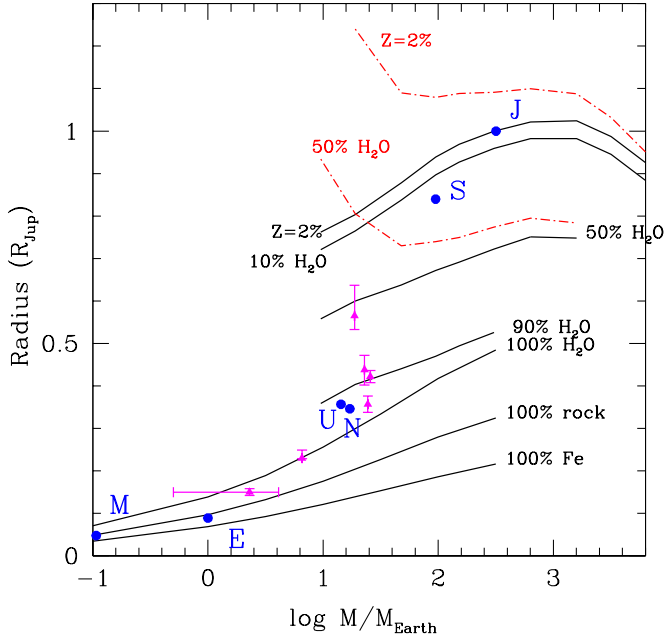}}
 \caption{Planetary radii at 4.5 Gyr as a function of mass, from 0.1 $\mearth$ to 20 $\mjup$. Models with solar metallicity (Z = 2\%) and with different amounts of heavy material (water, "rock" (i.e.  olivine or dunite), or iron) are shown (Baraffe et al. 2008, Fortney et al. 2007). Solid curves are for non-irradiated models while dash-dotted curves correspond to irradiated models at 0.045 AU from a Sun. The positions of Mars, the Earth, Uranus, Neptune, Saturn and Jupiter are indicated by solid points, while the most recent transiting Earth-like (Corot-Exo-7b, GJ1214b) and Neptune-mass (Hat-P-26b, GJ436b, Kepler-4b, Hat-P-11b) planets are indicated by solid triangles. 
 }
 \label{fig:MRth}
\end{figure}

Figure \ref{fig:MRth} focuses on the lowest mass part of the planetary domain, from 20 Jupiter masses down to Mars, i.e. going from gaseous giants to nearly incompressible matter. The figure displays the behaviour of the mass-radius relationship in this domain for various internal compositions, and highlights also the impact of stellar irradiation for a typical HD209458b-like system on the radius of a gas-dominated planet. Also indicated on the figure are the locations of the Solar System planets and of the recently discovered Earth-like\footnote{Note that Fig. 2 displays the revised 1$\sigma$ mass determination of Corot-7b (Pont et al. 2010).}, Super Earth or Neptune-like transiting objects.

\subsection{The planet radius anomaly}
\label{sec:anom}

On the other hand, as seen in Figure \ref{fig:MR}, a large number of transiting planets exhibit a radius significantly larger than predicted by the theory, even when including irradiation effects from the parent star. Denoting $R_{\mathrm{irrad}}$ such a theoretical radius, the
\textit{radius anomaly} of "Hot Jupiters" is thus defined as $(R_{\mathrm{obs}}-R_{\mathrm{irrad}})/R_{\mathrm{irrad}}$ (see e.g. Leconte et al. 2009, 2010c
or Fig. 10 of Baraffe et al. 2010). Several physical mechanisms have been suggested  to explain this radius anomaly. The most promising ones are discussed in details in Baraffe et al. (2010) and are quickly summarized below:

- tidal heating due to circularization of the orbit, as originally suggested by Bodenheimer et al. (2001). This suggestion has been revisited recently by Leconte et al. (2010a) using orbital equations which are valid at any order in eccentricity (Hut 1981, see also Eggleton et al. 1998). Indeed, all the previous calculations addressing this issue were based on a tidal model valid only for nearly circular orbits, as developed initially for our Solar System planets (Goldreich \& Soter 1966). As rigorously demonstrated in Leconte et al. (2010a) and Wisdom (2008), such a model severely underestimates the tidal dissipation rate as soon as the (present or initial) eccentricity is larger than about 0.2-0.3. Using tidal equations valid at any order in eccentricity shows that tidal dissipation, although providing a substantial source of energy and - for moderately bloated planets - leading to the appropriate radius, cannot explain the very bloated objects such as HD 209458b (Leconte et al. 2010a, Hansen 2010). It should be stressed that the aforementioned limitation of the so-called constant-$Q$ model does not have anything to do with the description of the dissipation mechanism in the star or the planet, as often misunderstood, but stems from the truncated expansion of the orbital equations. A dedicated discussion of these tidal effects is given in this volume by Leconte et al. 

- downward transport of kinetic energy originating from strong winds generated at the planet's surface by a small amount ($\sim 1\%$) of absorbed incident stellar radiation (Showman \& Guillot 2002). Although appealing, such a mechanism still needs to be correctly understood. Simulations par Burkert et al. 2005, for instance, do not produce such a dissipation (see Shownan et al. 2008 for a recent review). The identification of a robust mechanism for
transporting this energy deep enough is still lacking and an accurate (so far missing) description of the (small-scale) dissipative processes in such
natural heat engines is mandatory to assess the validity and the importance of this mechanism for hot-Jupiters
(see e.g. Goodman 2009). 

- ohmic dissipation in the ionized atmosphere of hot-Jupiters (Batygin \& Stevenson 2010). This scenario has received some support from recent 3D resistive MHD atmospheric circulation simulations of HD 209458b's weakly ionized atmosphere (Perna et al. 2010). According to these simulations, for
magnetic field strengths $B\gtrsim 10$ G, enough ohmic dissipation occurs at deep enough levels (from a few bars to several tens of bars)  to affect the internal adiabat and to slow down enough the planet's contraction to yield a significantly inflated radius. These results have to be confirmed by further studies, as quantifying the impact of non-ideal MHD terms and induced currents in numerical simulations is a challenging task.

- enhanced opacities ($\sim 10\times$ the solar mixture) in hot-Jupiter atmospheres, stalling the planet's cooling and contraction (Burrows et al. 2007). It should be stressed, however, that if the planet H/He envelope's global {\it metallicity} is enhanced at this level, the increased molecular weight will cancel or even dominate the opacity effect and will lead to a similar or smaller radius than the one obtained with solar metallicity (Guillot 2008). This scenario is thus so far an ad-hoc procedure and enhanced sources of opacities for a global solar-like metallicity must be identified, both theoretically and observationally.

- inefficient (layered or
oscillatory) convection in the planet's interior, due to a gradient of heavy elements either inherited from the formation stages or due to core erosion during the planet's evolution (Chabrier \& Baraffe 2007). Although layered convection is observed in many situations
in Earth lakes or oceans, due to the presence of salt concentrations (the so-called thermohaline convection), it remains unclear, however, whether this process can occur and persist under giant
planet interior conditions. 

In contrast to the first three scenarios, the last two ones (i) do not invoke an extra source of heating in the planet but rather an hampered output flux during the evolution, leading to a slower contraction rate, (ii) do not necessarily apply to short-period, irradiated planets only but could possibly also occur in planets at large orbital distances. 

None of these mechanisms has either been confirmed or ruled out so far. Note that they are not exclusive from each other and it might be possible that they all contribute, at some level, to the puzzling anomalously large radius problem.

\section{Departure of short-period planets from sphericity. Effect on the transit light curve and radius determination.}

Recent observations have shown that information about the departure of the planet from sphericity, due to rotationally or tidally induced forces, can be obtained from the analysis of planet transit light curves (Welsh et al. 2010, Carter \& Winn 2010a,b). Because of the tidal and/or rotational deformation (of both the planet and the star) the observed transit cross-section is smaller than the one corresponding to the genuine equilibrium radius of the planet (the one given by definition by 1D structure and evolution models). Recently, Leconte et al. (2010b) have investigated such a deformation of short-period planets and have shown that this deformation can have a non negligible impact on (i) the depth of the light curve itself, (ii) the radius of the planet inferred from this light curve. The impact on the depth of the transit is found to be of the order of a few percents for planets orbiting within about 0.04 AU from their host star, and can reach almost 10\% for the least massive short-period planets, such as e.g. WASP-19b or WASP-12b, leading to a $\sim 5\%$ effect on the planet's radius determination. These effects must be correctly taken into account when determining the proper equilibrium radius of the planet from the  transit observations, to be compared with the 1D theoretical models. As mentioned above, and demonstrated in Leconte et al. (2010b), the radius correction on the planet will always lead to a {\it larger} radius determination than the one obtained when ignoring aspherical deformation, therefore increasing the radius anomaly mentioned in \S \ref{sec:anom}. Leconte et al. (2010b) derive analytical expressions to take these deformation effects into account and to calculate the planet's proper triaxial shape (and thus proper equilibrium radius), for various relevant transiting planetary system conditions (see \S 5 of Leconte et al. 2010b). Using these analytical expressions, one can straightforwardly derive the correct transit depth and planet's radius from the observed (distorted) transiting object.

\section{The Brown Dwarf/Planet overlapping mass regime}
\label{sec:BDGP}

The distinction between BDs and giant planets has become these days a topic of intense debate. In 2003,
the IAU has adopted the deuterium-burning minimum mass, $\sim 10\,\mjup$,
 as the official distinction between the two types of objects.
We have already discussed the inadequacy of this limit in previous reviews (see e.g. Chabrier et al. 2007). As discussed in \ref{sec:formation} below, brown dwarfs and planets, although issued from
two different formation mechanisms, probably overlap in mass, so that there is no common mass limit between these two populations.
Therefore, the recent transit detection of massive companions in the substellar regime  ($ 5\, \mjup \lesssim M_p \lesssim M_{\mathrm{HBMM}}$) raises the questions about their very nature: planet or brown dwarf ?

As mentioned in \S 1, an internal heavy material enrichment yields a smaller radius, for a given mass, than a solar composition body. Therefore,
the m-R relationship provides in principle a powerful diagnostic to distinguish planets
from BDs in their overlapping mass domain. In practice, this diagnostic cannot always be obtained.
As shown in Leconte et al. (2009), for objects such as CoRoT-3\,b (Deleuil et al. 2008) or 
HAT-P-2\,b, with the revised radius determination (P{\'a}l  et al. 2010), the situation remains ambiguous. On one hand, the observations are consistent with these objects being irradiated solar-metallicity brown dwarfs. On the other hand, given the impossibility so far to assess the nature and, more importantly, the impact of the missing mechanism responsible for the anomalously large radius observed in some short-period planets (see \S \ref{sec:anom}), these objects can also be strongly inflated irradiated planets, with a substantial metal enrichment. As seen in Figure \ref{fig:MR}, several substellar objects in the mass range $\sim 3$-20 $ \,\mjup$ belong to this category, i.e. have a radius consistent with the object being either an irradiated brown dwarf or planet.
As discussed in Leconte et al. (2009),
this ambiguity can be resolved only in the case where the observed radius is significantly {\it smaller} than predicted for solar or nearly-solar metallicity (irradiated) objects. As mentioned above, this indeed reveals the presence of a significant global amount of heavy material in the transiting object's interior, as expected from planets formed by core accretion. This is, for instance, the case of Hat-P-20b (see Figure \ref{fig:MR}, and Fig. 2 of Leconte et al. 2010c), which is too dense to be a brown dwarf. According to the calculations of Leconte et al. (2010c), this object's radius determination implies more than 340 $\mearth$ of heavy material in the planet, i.e. a $Z\gtrsim 15\%$ global mass fraction. Such a heavy material mass fraction is compatible with, although at the upper end of, planet formation efficiency in protoplanetary disks (see eqn.(1) of Leconte et al. 2009), according to models of planet formation by core accretion (Mordasini et al. 2009). If the mass and radius of Hat-P-20b are confirmed, this object proves that planets can form up to at least 8$ \mjup$.

On the other hand,
the brown dwarf status of objects such as CoRoT-15\,b (Bouchy et al. 2010),
WASP-30\,b (Anderson et al. 2010) or LHS 6343 C (Johnson et al. 2010)
can not be questioned given their mass. Such masses can not be produced by the core accretion mechanism for planet formation, nor by gravitational instability in a disk at this orbit. As mentioned in \S1, the radius determination of these objects (at least the two last ones, given the large error bar for CoRoT 15\,b) confirms remarkably well the predicted m-R relationship in the BD domain. Comparison with this theoretical relation also shows that these objects are not significantly inflated, a consequence of the smaller incident flux contribution with respect to these object intrinsic internal energy compared with smaller objects.

\section{Constraints on planet formation mechanisms}
\label{sec:formation}

The observation of free floating objects with masses of the order of a few Jupiter masses in (low extinction) young clusters
(Caballero et al. 2007) shows that star and BD formation extends down to Jupiter-like masses,
with a limit set up most likely by the opacity-limited fragmentation, around a few Jupiter-masses (Boyd \& Whitworth 2005).
Observations show
that young brown dwarfs and stars share the same properties and are consistent with BDs and stars
sharing the same formation mechanism (Andersen et al. 2008, Joergens 2008; see Luhman et al. 2007 for a review), as supported by analytical theories of gravo-turbulent collapse of molecular clouds (Padoan \& Nordlund 2004, Hennebelle \& Chabrier 2008). On the other hand, the fundamentally different mass distributions of exoplanets detected by radial velocity surveys (Udry \& Santos 2007) clearly suggests a different formation
mechanism. The detection of transiting planets whose radius implies a large enrichment in heavy material, as mentioned in the previous sections, strongly supports the so-called core accretion scenario for planet formation (Pollack et al. 1996, Alibert et al. 2005, Mordasini et al. 2009). Conversely, this
large heavy material enrichment clearly excludes the
gravitational instability scenario (Boss 1997). The only remaining, although uncertain possibility for this latter is the formation of 
planets at very large distances ($\simgr 100$ AU), for the disk, assuming it is massive enough, to be cold enough
to violate the Toomre stability condition (Rafikov 2005, Whitworth \& Stamatellos 2006; see Dullemond et al. 2009 for a recent review on this issue).

According to these two different {\it dominant} formation mechanisms for stars/BDs and planets, these latter are supposed to have a substantial enrichment in heavy elements compared with their parent star, as observed for our own solar giant planets, whereas BDs of the same
mass, issued dominantly from the gravoturbulent collapse of a cloud, should have the same composition as their parent cloud, ie  a $Z\sim 2\%$ heavy element mass
fraction for a solar-like environment. Furthermore, the aforementioned brown dwarf detections down to a few ($\sim 5$) Jupiter masses, below the deuterium-burning limit, and the planetary nature of Hat-P-20b (if confirmed by further observations) are evidences that there is probably {\it no mass edge between planets and brown dwarfs but instead that these two populations of astrophysical bodies overlap}. 


\section{Conclusion and perspectives}

In this review, we have examined our present understanding and non-understanding of exoplanet formation, structure and evolution. The results can be summarized as follows:
\begin{itemize}
\item the theoretical mass-radius relationship in the brown dwarf and planetary regime seems to be confirmed by recent radius determinations of transiting massive brown dwarfs. When the object's radius is smaller than the one predicted for a gaseous body with solar composition, this m-R relationship enables us to distinguish planets from brown dwarfs in their overlapping mass domain and thus provides a key diagnostic to identify these two distinct populations. In other cases, the diagnostic remains ambiguous and the very nature of the transiting object can not be determined.
\item Present models of planet interior structure and evolution stand on relatively robust grounds and enable us to infer with reasonable confidence the gross internal structure and composition of these objects. Uncertainties in the EOS of various elements under the relevant conditions, however, prevent a detailed determination of this composition.
\item a large fraction of {\it gas dominated} transiting planets still exhibit a radius significantly larger than predicted by the models. Several physical mechanisms have been proposed to solve this "radius anomaly" problem but, so far, no firm conclusion about which one, if any, of these mechanisms is the correct one has been reached.
\item tidal energy dissipation due to circularization of the orbit, in the planet's interior, although providing a significant extra source of energy to the planet, has been shown not to be sufficient to explain the radius of the most bloated planets, including HD 209458-b. Indeed, when properly calculating the orbital evolution equations in case of a finite present or initial eccentricity, tidal dissipation is shown to occur too quickly during the planet's evolution to explain its present radius. Although a proper treatment of the contribution of dynamical tides is presently lacking, the equilibrium tide contribution calculated with the complete tidal equations still provides a lower limit for tidal dissipation and must be correctly calculated. Interestingly enough, recent observations of spin-orbit misalignment for planets orbiting F stars seem to point to a tidal dissipation in the star, and thus to a dynamical evolution of the system, which depends on the stellar mass, more precisely on the size of the stellar outer convection zone (Winn et al. 2010).
\item an update of the presently discovered transiting systems confirms the previous analysis of Levrard et al. (2009). Only a handful of these systems have enough total angular momentum to reach an orbital equilibrium state. For the vast majority of these systems, the planet experiences ongoing orbital decay and will eventually merge with the star, with the  dynamical  evolution timescale for the orbit semimajor axis and the stellar spin and obliquity being essentially the lifetime of the system itself (Levrard et al. 2009, Matsumura et al. 2010).
\item departure of both the parent star and the transiting planet from sphericity, because of either rotational or tidal forces, affects both the depth of the transit light curve and the planet's radius determination, and leads to an {\it underestimate} of this latter. This bias must be corrected to get a proper determination of the planet's genuine equilibrium radius, the one calculated with 1D structure models.
\item observations of Hat-P-20-b, if confirmed, show that planets form up to at leat about 8 $\mjup$ and thus the brown dwarf and planet mass regimes very likely overlap. Therefore, there is no common mass limit between these two populations of astrophysical bodies, stressing again the inadequacy of the definition put forward by the IAU.
\item the large number of transiting planets whose radius implies a substantial fraction of heavy material strongly supports the core accretion scenario formation for planets. In this scenario, the planet embryo originates from accretion of solids onto a core in the protoplanetary disk, leading eventually to dynamical accretion of gas dominated material above about 10 $\mearth$. Conversely, this same large metal enrichment excludes the gravitational instability scenario as the dominant formation mechanism for planets. 
\end{itemize}


\acknowledgments{The research leading to these results has received funding from the European Research Council under the European Community's 7th Framework Programme (FP7/2007-2013 Grant Agreement no. 247060).}

\end{document}